\documentclass[oneside,twocolumn,pra,amsmath,amssymb,superscriptaddress]{revtex4-1}

\usepackage{cases}
\usepackage{amsmath}
\usepackage{amssymb}
\usepackage{amsfonts}
\usepackage{amssymb}
\usepackage{dcolumn}
\usepackage{bm}
\usepackage{bbm}
\usepackage{graphicx}
\usepackage{xcolor}
\usepackage{array}
\usepackage{subfigure}
\usepackage{hyperref}

\hypersetup{colorlinks=true,
            breaklinks=true,
            pdfstartview=Fit,
            linkcolor=blue,
            citecolor=blue,
            urlcolor=blue}

\begin{document}
\title{Magnetic Gradiometer for Detection of\\ Zero- and Ultralow-Field Nuclear Magnetic Resonance}

\date{\today}

\author{Min Jiang}
\affiliation{
CAS Key Laboratory of Microscale Magnetic Resonance and Department of Modern Physics, University of Science and Technology of China, Hefei, Anhui 230026, China}
\affiliation{
Helmholtz Institute, Johannes Gutenberg University, 55128 Mainz, Germany}
\affiliation{
Synergetic Innovation Center of Quantum Information and Quantum Physics, University of Science and Technology of China, Hefei, Anhui 230026, China}

\author{Rom$\acute{\textrm{a}}$n Picazo Frutos}
\affiliation{
Helmholtz Institute, Johannes Gutenberg University, 55128 Mainz, Germany}

\author{Teng Wu}
\email{teng@uni-mainz.de}
\affiliation{
Helmholtz Institute, Johannes Gutenberg University, 55128 Mainz, Germany}

\author{John W. Blanchard}
\affiliation{
Helmholtz Institute, Johannes Gutenberg University, 55128 Mainz, Germany}

\author{Xinhua Peng}
\email{xhpeng@ustc.edu.cn}
\affiliation{
CAS Key Laboratory of Microscale Magnetic Resonance and Department of Modern Physics, University of Science and Technology of China, Hefei, Anhui 230026, China}
\affiliation{
Synergetic Innovation Center of Quantum Information and Quantum Physics, University of Science and Technology of China, Hefei, Anhui 230026, China}
\affiliation{
Synergetic Innovation Center for Quantum Effects and Applications, Hunan Normal University, Changsha, Hunan 410081, China}

\author{Dmitry Budker}
\affiliation{
Helmholtz Institute, Johannes Gutenberg University, 55128 Mainz, Germany}
\affiliation{
Department of Physics, University of California at Berkeley, California 94720-7300, USA}

\begin{abstract}

Magnetic sensors are important for detecting nuclear magnetization signals in nuclear magnetic resonance (NMR).
As a complementary analysis tool to conventional high-field NMR, zero- and ultralow-field (ZULF) NMR detects nuclear magnetization signals in the sub-microtesla regime.
Current ZULF NMR systems are always equipped with high-quality magnetic shieldings to ensure that ambient magnetic field noise does not dwarf the magnetization signal. 
An alternative approach is to separate the magnetization signal from the noise based on their differing spatial profiles, as can be achieved using a magnetic gradiometer.
Here, we present a gradiometric ZULF NMR spectrometer with a magnetic gradient noise of 17$~\textrm{fT}_{\textrm{rms}}~\textrm{cm}^{-1} ~ \textrm{Hz}^{-1/2}$ in the frequency range of 100-400 Hz, based on a single vapor cell (0.7$\times$0.7$\times$1.0 $\textrm{cm}^3$).
With applied white magnetic-field noise, we show that the gradiometric spectrometer achieves 13-fold enhancement in the signal-to-noise ratio (SNR) compared to the single-channel configuration.
By reducing the influence of common-mode magnetic noise, this work enables the use of compact and low-cost magnetic shields.
Gradiometric detection may also prove to be beneficial for eliminating systematic errors in ZULF-NMR experiments searching for exotic spin-dependent interactions and molecular parity violation.
\end{abstract}

\maketitle

\setlength{\parskip}{0em}

\section{Introduction}
Nuclear magnetic resonance (NMR), conventionally operated in large magnetic fields ($\sim$~T), is a powerful analytical technique in chemistry, biology, and medicine~\cite{Ernst1987,ZPLiang1999,KW1986}.
Zero- and ultralow-field (ZULF) NMR~\cite{Weitekamp1983, Lee1987, Theis2011, Blanchard2015, BlanchardD2016, TaylerMC2017} offers improved spectral resolution and untruncated spin interactions.
Combining with recently developed quantum-control techniques~\cite{Sjolander2017,Tayler2016, Sjolander20172,Bian2017, Jiang2017},
ZULF $\textrm{NMR}$ serves as a complementary tool to conventional high-field NMR.
For example, the absence of a large applied magnetic field allows for the measurement of antisymmetric spin-spin couplings, which are related to chirality~\cite{King12017} and have been proposed as a means for detecting molecular parity nonconservation~\cite{King22017}.
Furthermore, ZULF NMR has been recently applied to searches for axion and axion-like-particle dark matter~\cite{Garcon2017} and the nuclear spin-gravity coupling \cite{Teng2018}.

ZULF NMR typically involves low frequencies, so non-inductive sensors are necessary for detection.
Early ZULF NMR systems used superconducting quantum interference devices (SQUIDs) as magnetic field sensors~\cite{Storm2017, Greenberg1998, McDermott2002, Burghoff2005}.
One drawback is that SQUIDs must operate under cryogenic conditions.
Recent years have seen increased developments in atomic magnetometers~\cite{Budker2007}, both in sensitivity and portability.
A spin-exchange relaxation-free (SERF) atomic magnetometer~\cite{Allred2002, Kominis2003, Dang2010}, with a measurement volume of $0.45~\textrm{cm}^3$,
has a demonstrated sensitivity of $0.16~\textrm{fT}~{\rm{Hz}}^{-1/2}$, comparable to the most advanced SQUIDs.
SERF atomic magnetometers have been recently used in ZULF NMR to detect pure $\textrm{$J$-coupling}$ spectra at zero-field \cite{Ledbetter2009, Blanchard2013} and determine spin-coupling topology at near-zero magnetic field \cite{Appelt2010, Ledbetter2011}.

In ZULF NMR, the noise level of a measurement is frequently dominated by the ambient magnetic field noise.
Although such noise can be suppressed with magnetic shields, there is additional magnetic field noise due to Johnson currents from the shields themselves~\cite{Nenonen1996}.
There is also magnetic field noise from heaters, thermistors, and the magnetic nulling coils.
To reduce this noise, we use a gradiometric $\textrm{NMR}$ spectrometer which is based on a two-channel SERF atomic magnetometer.
Important steps towards multi-channel magnetometers have been shown~\cite{Kominis2003, Dang2010, Wyllie2012, Fang2014, Sheng2017}.
Compared with previous ZULF-NMR spectrometers~\cite{Ledbetter2009, Ledbetter2011, Blanchard2013, Theis2011, TaylerMC2017}, our gradiometric spectrometer is sensitive to the magnetic field gradient produced by the $\textrm{NMR}$ sample but insensitive to homogeneous magnetic fields.
The magnetic-field gradient noise level is measured to be 17$~\textrm{fT}_{\textrm{rms}}~\textrm{cm}^{-1} ~ \textrm{Hz}^{-1/2}$, which is less than the typical magnetic field gradient produced by NMR samples, e.g., 4$~\textrm{pT}_{\textrm{rms}}~\textrm{cm}^{-1} ~ \textrm{Hz}^{-1/2}$ for $^{13}$C-formic acid in our current setup.
By using the gradiometric spectrometer, we demonstrate signal-to-noise ratio (SNR)-enhanced measurement of liquid-state NMR samples under application of spatially homogeneous white magnetic field noise, 
with noise spectral density $\sim 0.3~ \textrm{pT}_{\textrm{rms}}~\textrm{Hz}^{-1/2}$, comparable to that in an unshielded environment~\cite{Bevilacqua2016, Seltzer2004}.

\begin{figure}[t]
\centering
\includegraphics[width=0.85\columnwidth]{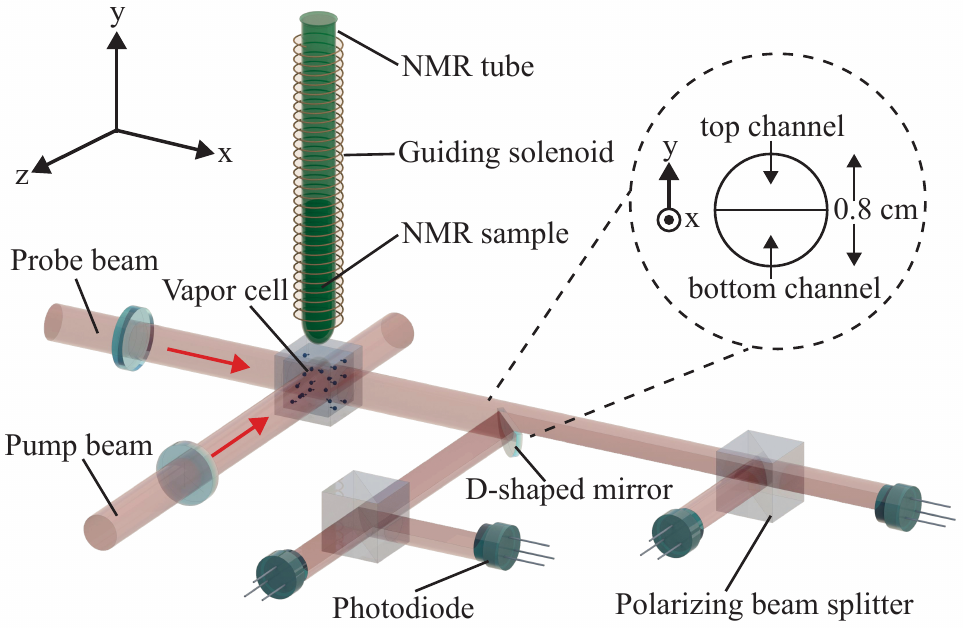}
\caption{Schematics of the gradiometric NMR spectrometer. The $^{87}$Rb vapor cell is resistively heated to 180$^\circ$C. A four-layer magnetic shield (not shown) isolates the vapor cell from external magnetic fields. Inset shows that the probe beam (with diameter $\sim 0.8~\textrm{cm}$) is split into top and bottom channels using a D-shaped mirror.}
\label{fig0}
\end{figure}

\section{Experiment}

The apparatus of the gradiometric spectrometer is shown in Fig.~\ref{fig0}.
The sensor consists of a ($^{87}$Rb) vapor cell, with an outer dimension of 0.7$\times$0.7$\times$1.0 $\textrm{cm}^3$ and a wall thickness of 1 mm.
The cell is optically pumped with a circularly polarized laser beam propagating in the -$z$ direction.
The laser frequency is tuned to the center of the buffer-gas (500 torr N$_{2}$) broadened and shifted D1 line of $^{87}$Rb.
The magnetic field is measured via optical rotation of linearly polarized probe laser light at the D2 transition propagating in the $x$ direction. 
The probe beam is expanded to have a diameter of 0.8 cm, which is able to cover the inner area of the vapor cell.
After the cell, the probe beam is split into two channels, top and bottom channels, respectively, with a sensing volume of  $0.1~\textrm{cm}^3$ for each channel, using a D-shaped mirror.
Gradiometric measurements are performed by taking the difference between the top- and bottom-channel signals.
Each channel operates as a SERF \cite{Allred2002, Happer1973} magnetometer, as described in \cite{TaylerMC2017}.
The inclusion of the buffer gas sets the diffusion length of rubidium atoms in one relaxation time to be about $(DT_2)^{1/2}\approx 0.01~\textrm{cm}$ ($D$ is the diffusion constant~\cite{Franz1976} and $T_2$ is the atomic spin relaxation time), which is much smaller than the spatial separation of the two channels ($\sim0.4~\textrm{cm}$).
As a result, each channel provides a local measurement of the magnetic field.
It is worth noting that the rubidium atoms in both channels have the same buffer-gas broadened and shifted resonance profile.
This is beneficial for cancelling some common-mode technical noise such as the intensity noise of the pump beam.

\begin{figure}[t]
\centering
\includegraphics[width=0.85\columnwidth]{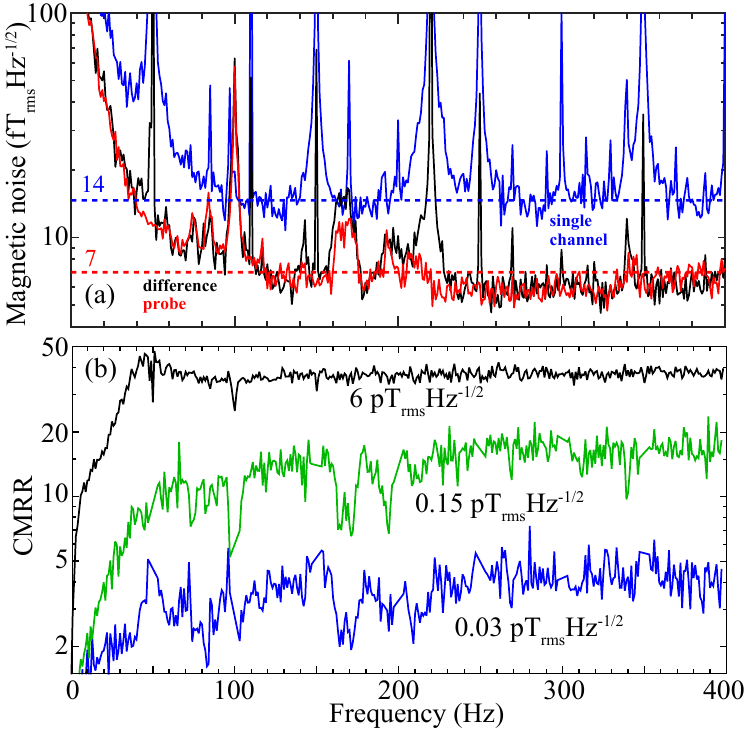}
\caption{Magnetic field noise and common-mode noise rejection ratio (CMRR) of the $\textrm{gradiometric}$ NMR spectrometer.~(a)~Magnetic field noise for the single channel $\sim14\ \textrm{fT}_{\textrm{rms}}~\textrm{Hz}^{-1/2}$ (blue line) and the difference of the two channels (black line) $\sim7\ \textrm{fT}_{\textrm{rms}}~\textrm{Hz}^{-1/2}$. The probe-beam noise is measured with the pump beam blocked and is shown with the red line $\sim7\ \textrm{fT}_{\textrm{rms}}~\textrm{Hz}^{-1/2}$.~(b) The CMRR for the gradiometric $\textrm{NMR}$ spectrometer under the application of white magnetic noise with different amplitudes. In the frequency range between zero and 40 Hz, the noise is mainly dominated by non-common-mode noise, which deteriorates the CMRR.}
\label{fig1}
\end{figure}

\section{Results and discussion}
Figure~\ref{fig1}{\color{blue}{(a)}} shows the magnetic field noise of the apparatus.
A single channel has a noise level of about $14~\textrm{fT}_{\textrm{rms}}~\textrm{Hz}^{-1/2}$, which is dominated by Johnson noise of the shield (Twinleaf MS-1) and is estimated to be $\sim 13~\textrm{fT}_{\textrm{rms}}~\textrm{Hz}^{-1/2}$ for the shield with an inner diameter of $14.7~\textrm{cm}$ \cite{Nenonen1996}. 
We perform the gradiometric measurement by taking the difference between top- and bottom-channel signals after calibrating the phase and amplitude of the channels (see Supplementary Information~\cite{SI}).
The subtraction of top and bottom channels gives a noise floor of $\sim 7~\textrm{fT}_{\textrm{rms}}~\textrm{Hz}^{-1/2}$, which is dominated by the intensity noise of the proble beam.
Dividing the noise floor by the spatial separation between the two channels gives the magnetic gradient noise of the gradiometric spectrometer, which is $\sim 17~\textrm{fT}_{\textrm{rms}}~\textrm{cm}^{-1} ~ \textrm{Hz}^{-1/2}$ from 100 to 400~$\textrm{Hz}$.

The common-mode noise rejection ratio (CMRR), which represents the ability to suppress common-mode magnetic field, is measured by applying a white magnetic field noise to our magnetometer through low-inductance Helmholtz coils (along the $y$ direction) with a diameter of $\approx 7~\textrm{cm}$.
The white magnetic field noise is generated from a function generator (DS345, Standford Research Systems) and is calibrated based on the response of the Helmholtz coils.
Since the noise floor of the difference of the two channels is no longer dominated by common-mode magnetic field noise, $\textrm{Figure}$~\ref{fig1}{\color{blue}(b)} shows that the CMRR depends on the amplitude of the applied white noise. 
For a white noise with an amplitude spectral density of 6~$\textrm{pT}_{\textrm{rms}}~\textrm{Hz}^{-1/2}$, the CMRR reached the maximum measured value, below 40 in a frequency range between 0 and 40 Hz, and
about 40 in a frequency range between 40 Hz and 400 Hz.
A CMRR larger than $100$ can be achieved by closed-loop operation~\cite{Sheng2017}, in which the signal is more robust against changes in sensor parameters than in the case of open-loop operations.

\begin{figure}[t]
\centering
\includegraphics[width=0.85\columnwidth]{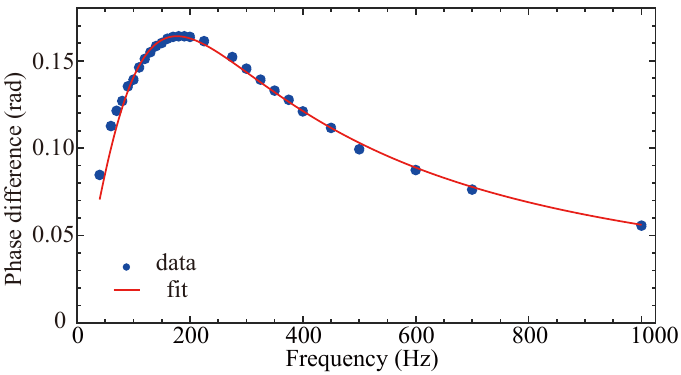}
\caption{The dependence of the phase difference between top and bottom channels on frequency. The points (circle) are experimentally measured phase differences through applying an oscillating $y$-field at different frequencies $f$. The experimental results are fitted to $\textrm{arctan} \textrm{[}\frac{f(A- B)}{f^2+A\cdot B} \textrm{]}$. Here, $A$ and $B$ are the bandwidths for the top and bottom channels, respectively. The fit (red line) indicates $A=206.9~\textrm{Hz}$ and $B=148.9~\textrm{Hz}$.}
\label{fig5}
\end{figure}

In order to improve the ability of the gradiometric $\textrm{NMR}$ spectrometer to suppress the common-mode magnetic field noise, it is necessary to calibrate and zero the phase difference between the two channels.
The procedure for phase-difference calibration is described in detail in the Supplemental Material~\cite{SI}.
The phase difference originates from the different dynamic response of the two channels and could be measured by applying an oscillating $y$-magnetic field at different frequencies.
$\textrm{Fig.}$~\ref{fig5} shows that the phase difference between the two channels is a function of frequency $f$, and can be fitted with $\textrm{arctan}\textrm{[}\frac{f(A- B)}{f^2+A\cdot B} \textrm{]}$.
Here, $A$ and $B$ are the bandwidths for the top and bottom channels, respectively.
The fit indicates that $A=206.9~\textrm{Hz}$ and $B=148.9~\textrm{Hz}$.
The difference in the bandwidth is due to the unevenly distributed power of the pump beam in the two channels.
The maximum phase difference $\approx 0.16~\textrm{rad}$ happens when the frequency is $f=\sqrt{A\cdot B}\approx 175.5$~Hz.
Based on these parameters, we can zero the phase difference between the two channels at different frequencies.

\begin{figure}
\centering
\includegraphics[width=1\columnwidth]{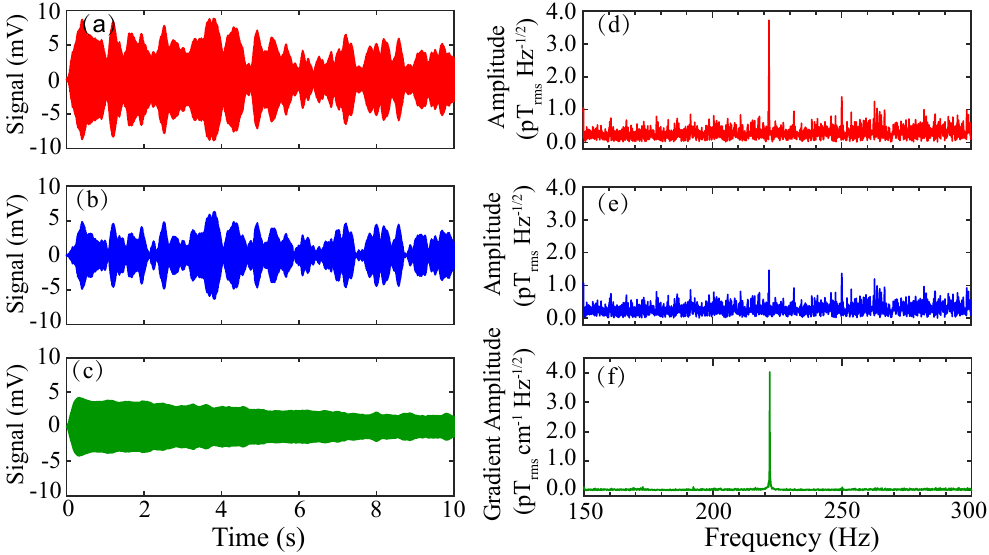}
\caption{Single-shot zero-field NMR of $^{13}$C-formic acid. White magnetic-field noise is applied though Helmholtz coils along the $y$-direction. The $^{13}$C-formic acid time-domain signals are observed in top (a), bottom (b), and gradiometric (c) channels at zero magnetic field, which are filtered with a second-order band-pass filter with cutoff frequencies at 220 Hz and 224 Hz in order to show the oscillating signals clearly. (d), (e) and (f) show the frequency-domain signals, which are obtained from the fast Fourier transform (FFT) of the unfiltered original time-domain signals from the three channels.}	
\label{fig2}
\end{figure}

The gradiometric spectrometer is used to detect ZULF-NMR signals.
NMR samples (typically $\approx 200 ~\mu \textrm{L}$) in standard 5-mm glass NMR tubes are polarized in a Halbach magnet (polarization field $B_p \approx 2.0~\textrm{T}$), after which the samples are shuttled into the magnetically shielded region, such that the bottom of the sample tube is $\sim0.3~\textrm{cm}$ above the $^{87}$Rb vapor cell.
A guiding magnetic field ($\sim 3\times 10^{-4}$~T) is applied during the transfer,
and is turned off within 10 $\mu$s prior to signal acquisition.
For a liquid-state $n$-spin-$\frac{1}{2}$ sample, the initial state~\cite{BlanchardD2016, TaylerMC2017} is $\rho_0=1/2^n(\mathbbm{1}+\sum_j \epsilon_j I_{jy})$ with $\epsilon_j=\gamma_j B_p/k_B T \sim 10^{-5}$, $\gamma_j$ denotes the gyromagnetic ratio of the $i$th spin, $k_B$ is the Boltzmann constant, $T$ is the temperature of the sample, and $\mathbf{I}_j=(I_{jx},I_{jy},I_{jz})$ is the spin angular momentum operator of the $i$th spin.
Approximating the NMR sample as a sphere, the magnetic moment due to the sample magnetization is $\mathbf{m}=(4\pi r_0^3/3)\mathbf{M}$, where $r_0$ is the radius of the sample, and $\mathbf{M}$ is the magnetization.
The magnetic field originated from the sample at the location of the sensor is 
\begin{align}
\mathbf{B}_s=\frac{\mu_{0}}{4\pi}\frac{3\hat{\mathbf{n}}(\mathbf{m}\cdot \hat{\mathbf{n}})-\mathbf{m}}{r^3}, 
\end{align}
where $\mu_0$ is the permeability of vacuum, $\hat{\mathbf{n}}$ is the unit vector pointing from the sample to the sensor and $r$ is the distance between them.
In our case, $\hat{\mathbf{n}}=-\hat{y}$.
Our gradiometric $\textrm{NMR}$ spectrometer is sensitive to magnetic field along $y$-axis, thus the measured signal from the nuclear magnetization has the form of 
\begin{align}
\mathbf{M}_y\propto \textrm{Tr}\{\rho(t)\sum_j I_{jy}\gamma_j\}.
\end{align}
In our setup, $r_0=0.21~\textrm{cm}$ and $r=0.85$ and $1.25~\textrm{cm}$ for the top and bottom channels, respectively.
Notably, there exists up to 50\% loss of NMR signals due to the gradiometric configuration.
This can be circumvented by slightly increasing the spatial separation between the two channels.
For example, the loss in the $\textrm{NMR}$ signal can be reduced to $10\%$ when the separation is increased to $1.0~\textrm{cm}$.

\begin{figure}[t]
\centering
\includegraphics[width=1\columnwidth]{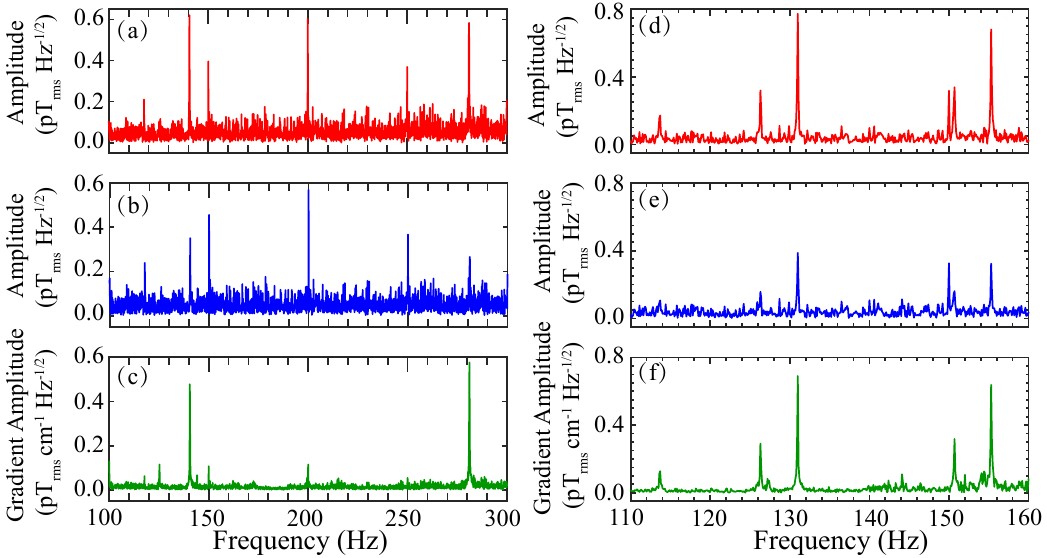}
\caption{Zero-field NMR spectra of $^{13}$C-methanol (a-c) and a portion of the spectra of fully labelled acetonitrile (from 110 to 160 Hz) (d-f). Each spectrum is the averaged result from eight transients. White magnetic-field noise is applied though Helmholtz coils along the $y$-direction. At zero magnetic field, the spectra of $^{13}$C-methanol ($^{13}$CH$_3$OH) are observed in top (a), bottom (b), and gradiometric (c) channels, respectively. The portion of the spectra of fully labelled acetonitrile ($^{13}$CH$_3$$^{13}$C$^{15}$N) are observed in top (d), bottom (e), and gradiometric (f) channels, respectively. Both (c) and (f) show that the gradiometer channel improves the SNR and reduces the spurious lines (the harmonics of 50 Hz). The full spectrum of $^{13}$CH$_3$$^{13}$C$^{15}$N is shown in the Supplemental Material~\cite{SI}. }
\label{fig3}
\end{figure}

With the gradiometric NMR spectrometer, high-SNR NMR spectroscopy in the low-frequency regime can be achieved even in a noisy environment.
ZULF-NMR spectroscopy of thermally polarized samples is discussed in Refs.~\cite{Appelt2010, Ledbetter2011}.
For a typical $\textrm{AX}_n$ system, where A and X are spin-$\frac{1}{2}$ particles, and each X couples to A with the same strength $J$, the resulting zero-field $J$-coupling spectra consist of a single line at $J$ for AX, a single line at $\frac{3}{2}J$ for AX$_2$ and two lines, one at $J$ and the other at $2J$, for AX$_3$ \cite{Theis2011}.
A single-shot NMR signal for $^{13}$C-formic acid (H$^{13}$COOH, from Sigma-Aldrich) obtained with our gradiometric NMR spectrometer is shown in Fig.~\ref{fig2}.
White magnetic field noise is applied though Helmholtz coils along the sensitive $y$ direction.
The amplitude spectral density of the noise is about $0.3~\textrm{pT}_{\textrm{rms}}~\textrm{Hz}^{-1/2}$.
A clear oscillating signal can be observed from the gradiometric channel (Fig.~\ref{fig2}{\color{blue}c}), compared with the signals in the single channels (Fig.~\ref{fig2}{\color{blue}a} and Fig.~\ref{fig2}{\color{blue}b}).
As shown in Fig.~\ref{fig2}{\color{blue}(d-f)}, the Fourier transform spectra consist of a single line at 221.9~Hz, and
the SNR for single shot are $\{15.1, 6.0, 202.9\}$ for the top, bottom, and gradiometric channels, respectively.
This indicates a 13-fold improvement in the SNR of the gradiometric channel compared with that of the single channel.
It is worth noting that the power-line-frequency noise (the harmonics of 50 Hz) are as well suppressed in the gradiometric channel.

We also use the gradiometric NMR spectrometer to measure molecules with more complex structures, as shown in Fig.~\ref{fig3}.
For the AX$_3$ case, the zero-field NMR spectra of methanol ($^{13}$CH$_3$OH, from Sigma-Aldrich) is shown in Fig.~\ref{fig3}{\color{blue}(a-c)}.
Each spectrum is the averaged result from eight transients.
Similarly, a white magnetic-field noise is applied into the gradiometric NMR spectrometer.
The observed SNR of the peak at $140.5$~Hz are $\{13.7, 7.8, 68.2\}$ and at $281.1$~Hz are $\{12.9, 5.9, 82.2\}$.
We present a simple rule to identify noisy peaks by comparing their gradient ratios with the expected value:
In Fig.~\ref{fig3}{\color{blue}(a)} and~\ref{fig3}{\color{blue}(b)}, a dubious line at $117.7$~Hz exists in both channels and the amplitude ratio between top- and bottom-channel signals is nearly one, which is sufficiently smaller than the expected signal amplitude ratio of the top and bottom channels.
This provides a good-confidence approach to distinguishing lines of spurious origin from $\textrm{NMR}$ signals.
Figures~\ref{fig3}{\color{blue}(d-f)} show the partial zero-field spectrum of fully labeled acetonitrile ($^{13}$CH$_3$$^{13}$C$^{15}$N, from Sigma-Aldrich) in the range of $110$~Hz to $160~\textrm{Hz}$.
The full spectrum of $^{13}$CH$_3$$^{13}$C$^{15}$N is presented in the Supplemental Material~\cite{SI}.
The SNR of a peak at $131.0$~Hz are $\{23.7,12.0,52.0\}$.
The spurious peak at $150~\textrm{Hz}$, which is close to one of the NMR peaks, is efficiently suppressed in the gradiometric channel.

\begin{figure}[t]
\centering
\includegraphics[width=1\columnwidth]{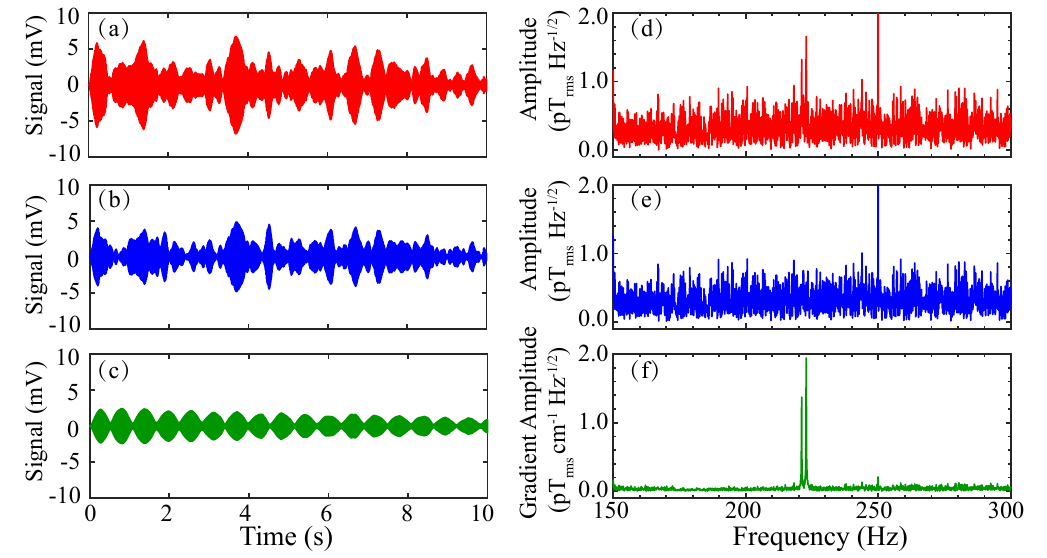}
\caption{Single-shot ultralow-field NMR of $^{13}$C-formic acid. White magnetic-field noise is applied though Helmholtz coils along the $y$-direction. The $^{13}$C-formic acid time-domain signals are observed in top (a), bottom (b), and gradiometric (c) channels in the presence of a small $z$ magneic field ($B_z \approx 31.90(18)$~nT), which are filtered with a second-order band-pass filter with cutoff frequencies at 220 Hz and 224 Hz in order to show the oscillating signals clearly. (d), (e) and (f) show the frequency-domain signals, which are obtained from the FFT of the unfiltered original time-domain signals from the three channels.}
\label{fig4}
\end{figure}

Ultralow-field NMR spectroscopy can be used for the identification of spin-coupling topology~\cite{Ledbetter2011, Appelt2010}.
In the presence of a small magnetic field $B_z$, the degenerate eigenstates of $^{13}$C-formic acid split into corresponding manifolds.
In Fig.~\ref{fig4}{\color{blue}(a)} and {\color{blue}(b)}, the single-shot time-domain NMR signals are perturbed by the ambient magnetic field noise.
However, we observe an obvious beating by using the gradiometric NMR spectrometer, as shown in Fig.~\ref{fig4}{\color{blue}(c)}.
In Fig.~\ref{fig4}{\color{blue}(f)}, a doublet with frequencies~\cite{Ledbetter2011} $J\pm B_z(\gamma_\textrm{H}+\gamma_\textrm{C})/2$ is visible with a higher SNR than those in Fig.~\ref{fig4}{\color{blue}(d)} and~\ref{fig4}{\color{blue}(e)}.
Here, $\gamma_\textrm{H}$ and $\gamma_\textrm{C}$ denote the gyromagnetic ratios for proton and carbon spins, respectively.
The frequency splitting of the doublet is measured to be $\Delta=B_z(\gamma_\textrm{H}+\gamma_\textrm{C})= 1.724(2)$~Hz and then $B_z$ is calculated to be $32.36(4)$~nT.
The doublet displays an asymmetry, i.e., the amplitude ratio between peaks at $J\pm B_z(\gamma_\textrm{H}+\gamma_\textrm{C})/2$ is about $1:0.7$.
In previous works on ultralow-field NMR using SERF atomic magnetometers \cite{Ledbetter2011, Ledbetter2012}, the asymmetry was also observed in the spectra. 
Evaluating high-order corrections to the eigenstates, the asymmetry of the doublet is found to be $\frac{2}{J}\Delta\approx 1.5\%$,
which is sufficiently smaller than the experimentally observed value ($\approx 35\%$).
We found that the asymmetry is due to changes in the sensitive directions in the SERF atomic magnetometers, i.e., the SERF magnetometer is not only sensitive to magnetic signals along the $y$ axis, but simultaneously sensitive to the signals along $x$ axis when a $z$-magnetic field is applied (see Supplemental Material~\cite{SI}).

\section{Conclusions and outlook}

We have experimentally demonstrated a gradiometric NMR spectrometer with a magnetic gradient noise of 17$~\textrm{fT}_{\textrm{rms}}~\textrm{cm}^{-1} ~ \textrm{Hz}^{-1/2}$ and a measurement volume for a single channel of $0.1~\textrm{cm}^3$.
With this device, we have demonstrated SNR-enhanced $\textrm{NMR}$ spectroscopy in the presence of environmental noise.
This opens the possibility of making a robust and portable NMR spectrometer, particularly in an unshielded environment where large common-mode magnetic field noise is introduced.
Our spectrometer is at an early stage of development, and many improvements are possible.
The amplitude of ZULF $\textrm{NMR}$ signals can be maximized by applying suitable magnetic field excitation pulses~\cite{BlanchardD2016, Ledbetter2009}.
With more optimization, such as decreased probe-beam fluctuations, uniform pump laser power and optimized spatial separation between the two channels, the gradiometric NMR spectrometer could be capable of detecting $\textrm{NMR}$ signals from samples with natural isotopic abundance for dilute nuclei, which are convenient for chemical analysis.
The gradiometric technique is not restricted just to the detection of NMR, but also opens up avenues of investigations of samples that generate magnetic field gradients, such as magnetic nanoparticles used in biomolecular labelling and cell separation~\cite{Yu2012, Bougas2018, Gleich2005}.
Moreover, recent theoretical work suggests that it is possible to measure molecular chirality and parity non-conservation effects in $\textrm{ZULF}$ $\textrm{NMR}$~\cite{King12017, King22017}.
Observation of such effects requires an oriented sample that can be obtained by applying a strong electric field, which creates unavoidable magnetic field noise~\cite{Romalis2001}.
An optimized gradiometric spectrometer is promising for sensing such chirality and parity non-conservation effects
while remaining robust against background magnetic field noise.

\section*{ACKNOWLEDGMENT}
We thank Jiankun Chen for drawing the setup diagram and Georgios Chatzidrosos for fruitful phase-related discussions.
This research was supported by the DFG (Deutsche Forschungsgemeinschaft) Koselleck Program and the Heising-Simons
and Simons Foundations, the European Research Council under the European Union’s Horizon 2020 Research, and Innovative Programme under grant agreement no. 695405 (T.W., J.W.B., and D.B.).
M. J. would like to acknowledge support from the China Scholarship Council (CSC) enabling his research at the Johannes Gutenberg-University Mainz. 
X. P. acknowledges the support from the National Key Research and Development Program of China
(2018YFA0306600), the National Key Basic Research Program of China (2014CB848700), National Natural Science Foundation of China (grants no. 11425523, 11375167, 11661161018, and 11227901), and Anhui Initiative in Quantum Information Technologies (grant no. AHY050000).

\end{document}